# Measurements of proton-induced radionuclide production cross sections to evaluate cosmic-ray activation of tellurium[#]

A. F. Barghouty[1], C. Brofferio[2,3], S. Capelli[2,3], M.Clemenza[2,3], O. Cremonesi[2,3], S.Cebrián[4], E. Fiorini[2,3,*] , R. C. Haight[5], E. B. Norman[6,7,8], M. Pavan[2,3] , E.Previtali[2,3] , B. J. Quiter[6], M.Sisti[2,3] , A. R. Smith[7], and S. A. Wender[5]

*[1] NASA-Marshall Space Flight Center, MSFC, AL 35812 U.S.A.*
*[2,3] Dipartimento di Fisica dell' Università di Milano-Bicocca and Istituto Nazionale di Fisica Nucleare Sezione di Milano-Bicocca, 20126 Milan, Italy*
*[4] Universidad de Zaragoza, 50009 Zaragoza, Spain*
*[5]Los Alamos National Laboratory, Los Alamos, NM 87545 U.S.A.*
*[6]Nuclear Engineering Department, University of California, Berkeley, CA 94720 U.S.A.*
*[7]Nuclear Science Division, Lawrence Berkeley National Laboratory, Berkeley, CA 94720 U.S.A.*
*[8]Physics Division, Lawrence Livermore National Laboratory, Livermore, CA 94551 U. S. A.*

[*]Corresponding author
E-mail address: ettore.fiorini@mib.infn.it
Tel.: 0039 02 64482424 FAX: 0039 02 64482463

## 1. INTRODUCTION

Experiments designed to study rare events such as the interactions of solar neutrinos [1], dark matter particles [2] or rare processes like double beta decay (DBD) [3] are carried out in underground laboratories. One of the main problems in such searches is the presence in various energy regions of background due to environmental radiation. The contribution due to cosmic rays [4] is strongly reduced by installing the experiment underground [5], sometimes with a further reduction by means of veto counters. While the local environmental radiation, mainly γ–rays or neutrons, can be suitably shielded with inert materials , special care should be devoted to reduce the intrinsic radioactivity of the detector itself or of the material immediately surrounding it [6]. In addition to the natural primordial radioactive contamination of detector material, one has to reduce the presence of radioactive nuclei produced by cosmic rays before the installation underground [7]. Shipping materials by airplane may be an issue due to the much higher cosmic ray flux at high altitude which could be two orders of magnitude higher.

The cosmic-ray contribution to the intrinsic radioactivity of detectors has been predicted by various authors using specific codes like COSMO [8] or general purpose codes like GEANT4 [9] .To evaluate the activation cross sections for particles at cosmic ray energies, a set of semi-empirical cross section formulas for proton-nucleus reactions developed by Silberberg and Tsao [10,11], and are used by codes like the mentioned COSMO and the recently introduced ACTIVIA [12]. To improve its reliability, the latter has been tuned to measurements of activation cross sections carried out with protons at various accelerators and can be applied to evaluate the effect of cosmic rays, because the fluxes [4] and cross-sections [12] of cosmic-ray neutrons in altitude can be assumed to be similar to those for protons [13, 14]. Many of these cross-section evaluations and measurements have been carried out for germanium [15-25] because of experiments with Ge diodes; others refer to different targets [26-34]. In this paper, we report extensive measurements of proton-induced radionuclide production cross sections on tellurium, for which limited data exist [12,35,36]. The main goal of this study is the identification of the radioactive isotopes that can be produced by cosmic ray exposure of $TeO_2$ cryogenic detectors (or bolometers) [37].

$TeO_2$ bolometers are used to search for neutrinoless DBD of $^{130}$Te which has a natural abundance of 33.8 % and a transition energy ($Q_{\beta\beta}$) of 2527.0 keV [38,39]. The best sensitivity on the absolute neutrino mass, comparable to the one of $^{76}$Ge and $^{100}$Mo experiments, has been reached with the CUORICINO experiment [40,41] based on an array of 62 crystals of $TeO_2$ with a total mass of about 40.7 kg. To improve this result a larger scale experiment is presently under construction : The Cryogenic Underground Observatory for Rare Events (CUORE) [42,43] will employ approximately 750 kg of $TeO_2$ bolometers in a search for the neutrinoless DBD of $^{130}$Te. Its aim is to reach a sensitivity on the absolute neutrino mass as indicated by the results of neutrino oscillations in the inverse hierarchy hypothesis. Minimizing the background in the vicinity of the $Q_{\beta\beta}$ value is crucial for the success of this experiment. Thanks to the deep underground location of the apparatus (that is being assembled in the underground Laboratori Nazionali del Gran Sasso, Italy) and to a careful design of the gamma-ray and neutron shields [44,45], the main background sources in CUORE will come from the radioactivity of the detector constructing materials [42,43,46,47]. Among these sources, cosmogenic activated isotopes may play a relevant role. For this reason a special protocol for the crystal procurement have been studied [46,47] in order to minimize the exposure of $TeO_2$ crystals to cosmic rays during growth and transportation: crystals are delivered from China (where they are grown) to Italy by ship in order to avoid the larger activation due by air transportation, then are immediately stored underground. The total exposure to sea-level cosmic rays does not exceed 45 days (starting from growth) while the storage underground is between 4 and 1 years long, depending on the production batch.

To demonstrate that this procurement protocol is able to ensure acceptable low levels of cosmogenic activation is not straightforward. Indeed, a direct evaluation of the activity of the potentially dangerous isotopes is often impossible: whatever the technique used the sensitivity is too low (as an example CUORE requirements for the $TeO_2$ crystal contamination in $^{60}Co$ are below one μBq/kg [42]).

For this reason the only solution is to rely on the computation of the expected activity on the basis of crystal exposure time, cosmic ray fluxes and activation cross sections.

## 2. EXPERIMENTAL SETUP

Exposures to the proton beams of $TeO_2$ samples have been performed in USA at the Los Alamos National Laboratory's Neutron Science Center at 0.8 GeV and in Europe at CERN at 1.4 and 23 GeV. The content of other elements in our samples was always less that a μg/g. Before exposure, each sample was tested for the presence of relevant activities produced by cosmic rays which were found to be negligible. All accelerator activation has been followed by γ-ray spectroscopic analysis of the irradiated samples at different times after irradiation.

### 2. 1 MEASUREMENTS CARRIED OUT IN USA

The 0.80-GeV irradiation was conducted in the "Blue Room" at the Los Alamos National Laboratory's Neutron Science Center (LANSCE). The Te target was fabricated from 99.9999% pure Te Alpha Aesar metallic powder that was pressed into a 6.9-mm thick disk that had 93.4% of normal Te metal density. To confirm that the proton beam was impinging entirely upon the target, a phosphor was placed in the target position before irradiation. A stack of targets was placed in air and irradiated with 800 ± 5 MeV protons for approximately 5 minutes. The total beam fluence was monitored by a beam current integrator in order to have approximately $10^{13}$ protons incident upon the target stack. The stack was made up of 6.4 mm thick polyethylene in front of the Te target and then an aluminium backing of 2 mm thickness. More precise proton fluence numbers were extracted from the production of $^{22}Na$ and $^{7}$Be in the monitor materials. The proton fluence deduced from the monitor foils was determined to be (1.2 ± 0.1) x $10^{13}$ p·cm $^{-2}$. The cross sections for production of these isotopes from $^{27}$Al have been previously determined using a very similar experimental setup at LANL [49], so these cross sections were used instead of evaluated data. The polyethylene monitor was analyzed using evaluated data for the C(p, x)$^{7}$Be cross section [50]. It should be noted that the energy lost by protons while traversing the target stack was significantly less than the overall uncertainty in the beam energy and therefore the proton energies are assumed to remain constant throughout the stack.

After the bombardment at 0.8 GeV at LANSCE, the samples were transported from Los Alamos back to Lawrence Berkeley National Laboratory (LBNL) where counting began approximately 15 days after irradiation. The samples were initially assayed with a 25% relative efficiency (with respect to that of a 7.62 x 7.62 cm NaI detector) n-type coaxial high-purity germanium detector. The samples were counted multiple times over the course of a week in order to extract half-life information from lines that could not be readily identified. These counting runs occurred at both 7.6 cm from the detector's end cap and with the sample against the end cap. Later, the targets were counted for longer periods with an 80% relative efficiency high purity germanium detector at LBNL's Low Background Facility (LBF). They were counted there intermittently for approximately one year to allow weaker long-lived lines to appear out of the Compton continuum. These counting runs occurred with samples either at 12 cm from the end cap or touching the end cap. All of the measurements of the targets irradiated at LANSCE were performed using ORTEC PC-based data acquisition systems. Figure 1 illustrates an energy spectrum of the Te target taken approximately 16 days after the LANSCE irradiation.

To determine the detection efficiencies for the four counting arrangements described above, calibrated point sources of $^{22}$Na, $^{54}$Mn, $^{57}$Co, $^{60}$Co, $^{109}$Cd, $^{137}$Cs, $^{152}$Eu and $^{228}$Th were used to establish a full-energy efficiency calibration curve of the detectors for point sources. The thick targets that were being counted, however, were sufficiently different from point sources to warrant further effort. In order to establish first-order corrections for both the geometric and self-attenuating effects of the thick targets, an identical un-irradiated Te target (henceforth referred to as “blank”) was placed at the described distance and the point source was placed behind the blank. The efficiency curves resulting from these two geometries were then averaged to produce the effective efficiency curve for a given target being counted by a given detector. While this method proved effective for the calibration of counting geometries where the targets were further from the detector, coincident summing of lines prevented it from being practical when the targets were placed directly against the end caps of the detectors. In this case we employed a method similar to that described by Gmuca and Ribansky [51]. An efficiency curve for a geometry with the source further from the detector is determined as above, then using only the sources with non-coincident lines, $^{57}$Co, $^{109}$Cd, $^{137}$Cs, and $^{54}$Mn, the ratio of efficiencies at the two geometries is determined and a curve is fit to that ratio. Then the longer-distance efficiency curve is multiplied by the aforementioned curve resulting in the final efficiency curve for these close geometries.

A brief inspection of the spectra shown in Fig. 1 allows one to gain an idea as to how many lines were present in the spectra. Each line was identified by energy and half-life. Where lines could be attributed to multiple isotopes, other lines were searched for to determine if one could positively identify which isotope was producing the line. If there was no way to confirm the presence of an isotope because it only had a single line or its other lines were also irresolvable, information from the half-life was employed. Cross sections were determined using the extracted γ-ray peak areas, measured γ-ray detector efficiencies, tabulated γ-ray intensities [51], measured target thickness, measured proton fluence and the start and stop times of the γ-ray counting.

## 2. 2 MEASUREMENTS CARRIED OUT IN EUROPE

Crystals of $TeO_2$ with natural isotopic composition of tellurium [48] were exposed in air to two proton beams of different energies at CERN. Details on the samples and on the exposures are reported in Table I. A single exposure of a telluride crystal of 0.457 g mass and irregular shape was performed at the 1.4 GeV beam of ISOLDE using the RABBIT system [52]. The cross section of the beam was approximately circular with a radius at half maximum of ~2 cm, thus totally covering the sample. The geometrical beam profile was taken into account to evaluate the effective proton fluence on the target. Due to the small thickness of the target, energy loss in it was negligible. The total beam fluence was monitored by a beam current integrator and known to within 10%.

Two exposures were also carried out in air with the 23 GeV beam at the IRRADIATION 1 Facility at the CERN PS East Hall [53,54]. The beam fluence was monitored with a current integrator and tested by activation of a 100 μm thick Al foil. The overall uncertainty on the fluence was 10%. The first exposure was carried out with a crystal of $TeO_2$ of irregular shape with a mass of 0.428 g and a low proton fluence in order to pre-evaluate the expected activity after irradiation. Also in this case the cross section of the beam did totally covered the sample and the beam profile was used to evaluate the effective proton fluence on the target. The loss of protons in the crystal is negligible.

A second exposure was performed later with a larger $TeO_2$ crystal of 23 x 22 mm sides and 1.6 mm thickness. As for the exposure at 1.4 GeV, the energy loss of the protons inside the target was negligible. The cross section of the beam was approximately circular with a radius at half maximum radius of ~ 8 mm. Since the beam did not cover the entire target the effective fluence through it was calculated taking into account, also in this case, the beam profile.

For radiation safety reasons and in order to avoid transportation from the irradiation zones, γ-ray spectroscopy for the CERN irradiated samples was first performed for period of a few months after exposure by the Irradiation Facility of CERN with standard CANBERRA Ge spectrometers with relative efficiency of up to 40% (with respect to that of a 7.62 x 7.62 cm NaI detector). However, since the presence of long-living isotopes is more relevant for a low-background underground experiments like CUORE, the sample irradiated at 1.4 GeV and the second one of those irradiated at 23 GeV were shipped to Milan years after the exposure, when the activity due to the short-living isotopes had sufficiently decreased.

The irradiated samples were then analyzed with two high purity germanium detectors, one of ~30% and the other one of ~60% relative efficiency. The energy resolutions of the two detectors at 1.33 MeV were 1.77 keV and 2.01 keV FWHM, respectively. All detectors were operated in the low-radioactivity laboratory of the Italian National Institute for Nuclear Physics (INFN) and of the University of Milano-Bicocca in the basement of the Territorial and Environmental Science Department.

Monte Carlo simulations were done in order to correctly evaluate the HPGe absolute detection efficiency ($\varepsilon_{abs}$) considering the relative position of the samples with respect to the detector. The efficiency of the detector for the analyzed radioisotopes was reconstructed with the aim of evaluate all possible aspects, especially those connected with coincidence summing effect that can influence the full energy peak efficiency for some radionuclides (for example: the complete decay scheme of $^{60}$Co, $^{65}$Zn, $^{54}$Mn, $^{102m}$Rh and $^{110m}$Ag were considered).

The software utilized named Jazzy, is based on GEANT4 [9]. This code has been tested by exposing our detectors to several certified gamma ray sources (point like and extended) of the Italian Authority ENEA and found to reproduce the entire spectrum (50 keV up to 2.5 MeV) within 5% error [6]. Taking into account the irregular shape of samples, it was assumed a systematic error of 10% in the overall gamma ray efficiency.

The spectra obtained with the γ-ray measurements performed 4.58 and 2.83 years after the exposures at 1.4 and 23 GeV are reported in Figures 2 and 3, respectively. They clearly show the presence of long-living isotopes produced by proton activation. Identifications of the major peaks produced by the irradiation are also indicated (Figures 2b and 3b).

## 3. RESULTS AND DISCUSSION

Our primary aim was to determine the production cross sections of long-lived isotopes, such as $^{60}$Co, and other long-living nuclei, whose presence could contribute to the background in the energy region of interest (2527 keV in the case of $^{130}$Te neutrinoless DBD). The quoted errors (90% c.l.) include in addition to statistics also the uncertainties on proton fluence and detector efficiency. The gamma rays produced by the decay of $^{60}$Co in the proton-irradiated Te targets are initially obscured by the activity of short-lived isotopes and can only be reliably measured months or years after the exposures. For completeness, however, we also report our results for relatively short-lived isotopes, which could play a role in the first period of the underground experiment, especially if the material was shipped by airplane.

The experimental results obtained from the irradiations performed at LANSCE and CERN are reported in Tables II and III, for long and short-living isotopes, respectively. We have calculated and also reported in the Tables the production cross sections calculated using the latest version of the semi-empirical formulae of Silberberg and Tsao (S&T) [10,11] (using the YIELDX routine). We note that the code only calculates cross sections for a specific daughter to be produced from a specific target. Thus, to determine each production cross section as defined above, calculations must be performed for each naturally occurring target isotope yielding the daughter nuclide as well as all short-lived precursors that decay to the daughter of interest. In the S&T calculations, no discrimination is made between meta-stable and ground states.

Before comparing our results with predictions, we would like to stress that the evaluation of Silberberg and Tsao is based on the hypothesis of activation by spallation [55]. This is definitely correct for product nuclei with atomic weight significantly different from those of the isotopes of Tellurium, but not when A is close to them. We found in fact an important presence of the isotopes $^{121}$Te, $^{121m}$Te, $^{123}$Te and $^{129}$Te due to the large cross section of thermal neutrons which were present during irradiation. A similar consideration applies to the production of $^{124}$Sb, $^{125}$Sb, and $^{126}$Sb where one can expect a large contribution from (n,p) reactions by fast neutrons on the $^{124}$Te, $^{125}$Te and $^{126}$Te nuclei which have a large isotopic abundance (4.816, 7.14 and 18.95% , respectively). For this reason we have not reported data on the cross section for the production of all these isotopes.

We would like to note that for two isotopes, $^{48}$V and $^{103}$Ru, there is a cumulative effect which could increase the estimated production. In addition to the direct production, $^{48}$V is produced by $\beta^+$/EC decays from $^{48}$Fe => $^{48}$Mn => $^{48}$Cr =>$^{48}$V while $^{103}$Ru could be produced by $\beta^-$ decays from $^{103}$Y => $^{103}$Zr => $^{103}$Nb => $^{103}$Mo => $^{103}$Tc => $^{103}$Ru. For this reason we have evaluated the expected direct activation cross sections of $^{48}$Fe, $^{48}$Mn, $^{48}$Cr and $^{48}$V and sum them to get cumulative cross section for $^{48}$V. The same has been done to evaluate the activation cross section for $^{103}$Ru. As can be seen from the Tables, the calculated and measured are typically within a factor 2 of each other, as was also reported in previous published paper on proton spallation reactions on Te [35,36]. Larger deviations from the calculated values and the experimentally measured ones are found for different nuclei and for different proton energies by Yu. T. Titarenko et al [29].

## 4. CONCLUSIONS

The above reported cross sections can be used to evaluate the contribution by cosmic ray activation of $TeO_2$ crystals to the background of CUORE. Limiting ourselves to long living nuclei (Table II), the only two isotopes able to contribute to the background in the Region Of Interest (ROI) for the $^{130}$Te Double Beta Decay study are $^{60}$Co and $^{110m}$Ag. Indeed all the others have a maximum energy release in the crystal far below the ROI. A rough evaluation of the initial activity of these isotopes in $TeO_2$ crystals is obtained using the cross sections reported in Table II and binning the cosmic ray spectrum at sea level accordingly. For an average exposure of 45 days, we obtain an activity of about $2 \times 10^{-8}$ Bq/kg in $^{60}$Co and $4 \times 10^{-6}$ Bq/kg in $^{110m}$Ag. These activities will decrease before the start of the experiment (the storage time varies between 5 and 1 years depending on the crystal production batch) by a factor 2 (for $^{60}$Co) and 9 (for $^{110m}$Ag) leading to a global contribution in the ROI that – according to preliminary evaluations - is at least one order of magnitude below the target counting rate for CUORE.

**ACKNOWLEDGMENTS**

We would like to express our gratitude to the ISOLDE Collaboration and in particular to Alexander Helert and Ulli Koester and to the ISOLDE Technical group for assistance and support. We are also grateful to the CERN IRRADIATION FACILITY and in particular to Maurice Glaser for help and advice in the exposures at 23 GeV. This work was supported in part by the U. S. Department of Energy under contract numbers DE-AC52-07NA27344 at LLNL and DE-AC02-05CH11231 at LBNL.

1 2 3 4 5 6 7 8 9 10 11 12 13 14 15 16 17 18 19 20 21 22 23 24 25 26 27 28 29 30 31 32 33 34 35 36 37 38 39 40 41 42 43 44 45 46 47 48 49 50 51 52 53 54 55 56 57 58 59 60 61 62 63 64 65

| Proton Energy (GeV) | Fluence (p/cm$^2$) | Mass $TeO_2$ (g) | Mass Te (g) | Delay Time (h) | Measurement Live Time (s) |
|---|---|---|---|---|---|
| 1.4 (a) | $1.5 \times 10^{15}$ | 0.457 | 0.365 | 1240.5 | 297022 |
| 1.4 (b) | $1.5 \times 10^{15}$ | 0.457 | 0.365 | 4827.5 | 251358 |
| 1.4 (c) | $1.5 \times 10^{15}$ | 0.457 | 0.365 | 40124.5 | 583178 |
| 23 (0) | $9.41 \times 10^{10}$ | 0.424 | 0.339 | 848.2 | 50400 |
| 23 (a) | $6.1 \times 10^{12}$ | 51.11 | 40.9 | 2224 | 1097 |
| 23 (b) | $6.1 \times 10^{12}$ | 51.11 | 40.9 | 2591 | 50400 |
| 23 (c) | $6.1 \times 10^{12}$ | 51.11 | 40.9 | 13606 | 200000 |
| 23 (d) | $6.1 \times 10^{12}$ | 51.11 | 40.9 | 24793 | 861228 |

**TABLE 1.** CERN Irradiations data and delay and live times of Gamma ray spectrometry measurements

| Isotope | Half Life (d) | Gamma Ray Energy (keV) | Decay Mode | σ Expt. 0.8 GeV (mb) | σ S&T 0.8 GeV (mb) | σ Expt. 1.4 GeV (mb) | σ S&T 1.4 GeV (mb) | σ Expt. 23 GeV (mb) | σ S&T 23 GeV (mb) |
|---|---|---|---|---|---|---|---|---|---|
| $^{54}Mn$ | 312 | 1377 | EC | 0.04±0.02 | 0.11 | | 0.40 | 1.5±0.3 | 4.0 |
| $^{57}Co$ | 272 | 836 | EC | 0.05±0.01 | 0.20 | 0.15±0.05 | 0.73 | 0.87±0.09 | 1.1 |
| $^{60}Co$ | 1923.6 | 2824 | $\beta^-$ | 0.09±0.04 | 0.22 | 0.20±0.04 | 0.77 | 0.75±0.08 | 1.15 |
| $^{65}Zn$ | 244 | 1352 | EC, $\beta^+$ | 0.11±0.02 | 0.46 | 1.2±0.3 | 1.5 | 1.8±0.2 | 2.4 |
| $^{75}Se$ | 120 | 864 | EC | 0.26±0.08 | 1.1 | 3.2±0.3 | 2.7 | 3.1±0.3 | 4.3 |
| $^{88}Y$ | 107 | 3623 | EC, $\beta^+$ | 3.1±1.2 | 1.3 | 4.6±0.8 | 5.0 | 4.2±0.6 | 3.1 |
| $^{102}Rh$ | 207 | 2323<br>1151 | EC, $\beta^+$<br>$\beta^-$ | 4.9±1.2 | 7.9 | 1.5±0.3 | 11 | | 5.8 |
| $^{102m}Rh$ | 1058.5 | 2323 | EC, $\beta^+$ | 4.0±0.6 | 1.3 | 2.4±0.5 | | 1.5±0.2 | |
| $^{110m}Ag$ | 250 | 2892 | $\beta^-$ | 24.6±3.7 | 16.0 | 1.9±0.3 | 1.2 | 0.88±0.59 | 0.64 |
| $^{113}Sn$ | 115 | 1036 | EC | 12.4±1.9 | 19.5 | 27±5 | 19 | | 11 |

**TABLE 2**. Results for activation of isotopes with half life > 100 days

| Isotope | Half Life (d) | Gamma Ray Energy (keV) | Decay Mode | σ Expt. 0.8 GeV (mb) | σ S&T 0.8 GeV (mb) | σ Expt. 1.4 GeV (mb) | σ S&T 1.4 GeV (mb) | σ Expt. 23 GeV (mb) | σ S&T 23 GeV (mb) |
|---|---|---|---|---|---|---|---|---|---|
| $^{46}Sc$ | 83.8 | 2367 | $\beta^-$ | 0.05±0.01 | 0.11 | | 0.29 | 1.5±0.3 | 2.9 |
| $^{48}V$ | 16 | 4012 | EC, $\beta^+$ | | | | 0.12 | 24±5 | 1.2 |
| $^{56}Co$ | 77.3 | 4566 | EC, $\beta^+$ | 0.10±0.02 | 0.18 | | 0.04 | 0.18±0.03 | 0.40 |
| $^{58}Co$ | 70.9 | 2307 | EC, $\beta^+$ | 0.04±0.01 | 0.38 | 0.5±0.1 | 1.45 | 1.7±0.3 | 2.2 |
| $^{59}Fe$ | 44.5 | 1565 | $\beta^-$ | | | 0.3±0.1 | 0.20 | 0.74±0.17 | 0.47 |
| $^{83}Rb$ | 86.2 | 909 | EC | 1.6±0.2 | 2.6 | | 4.1 | 5.3±0.8 | 3.8 |
| $^{84}Rb$ | 32.8 | 2681 | EC, $\beta^+$ | 0.40±0.06 | 0.4 | 4.8±0.8 | 1.7 | 4.2±0.7 | 1.6 |
| $^{85}Sr$ | 64.8 | 1065 | EC | 2.1±0.3 | 3.1 | | 6.1 | 5.4±0.9 | 4.8 |
| $^{88}Zr$ | 83.4 | 673 | EC | | | | 7.0 | 2.8±0.6 | 4.2 |
| $^{87}Y$ | 3.3 | 1862 | EC, $\beta^+$ | 2.9±0.4 | 3.2 | | | | |
| $^{95}Nb$ | 34.97 | 926 | $\beta^-$ | | | | 1.2 | 1.5±0.5 | 0.58 |
| $^{89}Zr$ | 3.3 | 2833 | EC, $\beta^+$ | 3.6±0.5 | 4.7 | | | | |
| $^{92m}Nb$ | 10.2 | 2006 | EC, $\beta^+$ | 0.23±0.03 | 3.6 | | | | |
| $^{95}Nb$ | 34.97 | 926 | $\beta^-$ | 0.85±0.17 | 0.5 | | | | |
| $^{95m}Tc$ | 61 | 1691 | EC, $\beta^+$ | 0.46±0.07 | 7.6 | | | | |
| $^{100}Pd$ | 3.6 | 361 | EC | 6.4±1.1 | 2.7 | | | | |
| $^{103}Ru$ | 39.3 | 763 | $\beta^-$ | 1.4±0.2 | 0.6 | | 0.82 | 19±3 | 0.37 |
| $^{105}Ag$ | 41.3 | 1345 | EC | 10.1±1.7 | 19.7 | | 15 | 3.7±1 | 9.3 |
| $^{111}In$ | 2.8 | 865 | EC | 16.4±3.0 | 30.4 | | | | |
| $^{114m}In$ | 49.5 | 1452 | EC, $\beta^+$ | 11.6±2.3 | 3.7 | 9.2±0.9 | 3.1 | 8.0±1,2 | 1.9 |

**TABLE 3**. Results for activation of isotopes with half life < 100 days

# FIGURES

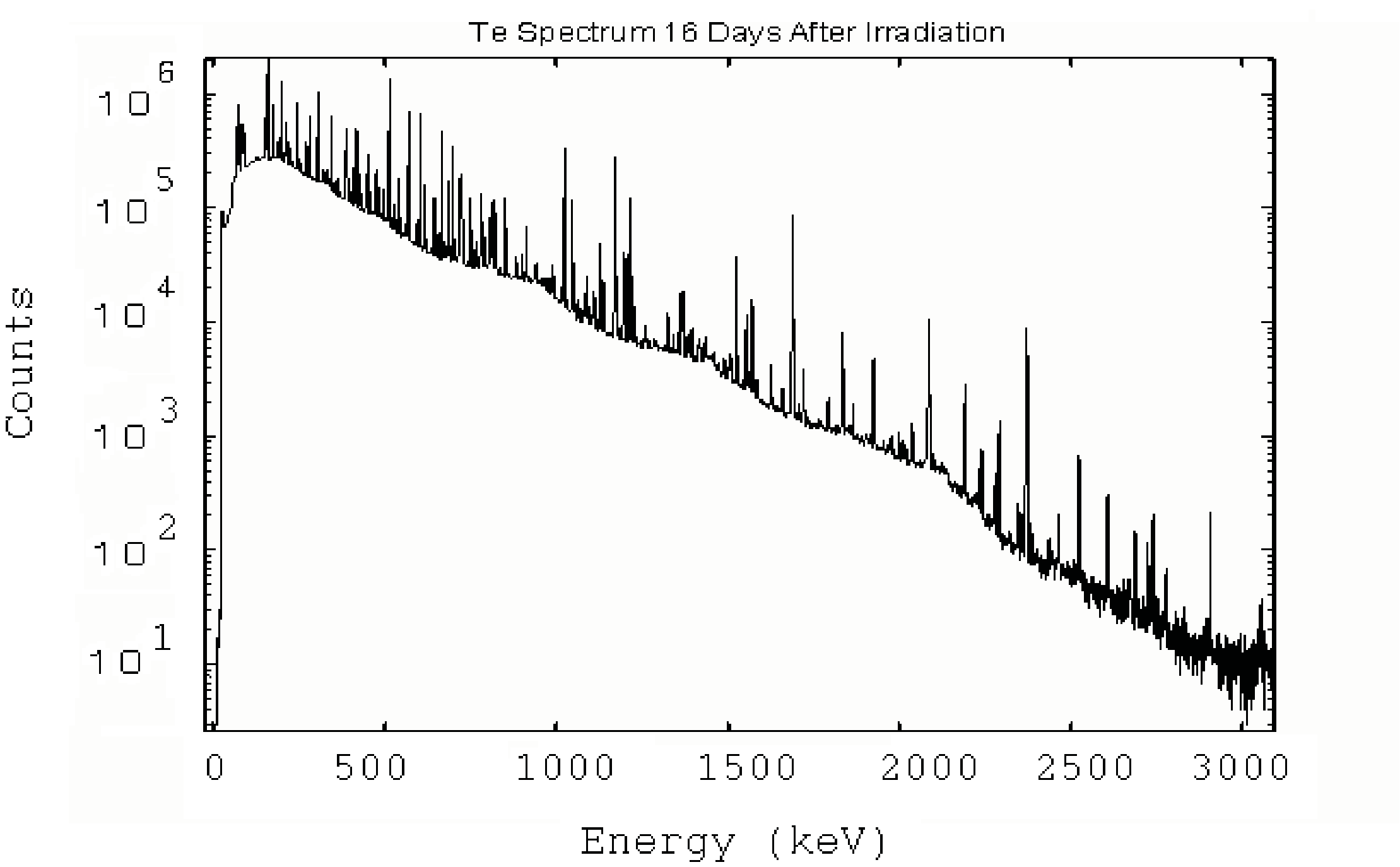

**FIGURE 1:** $\gamma$-ray spectrum recorded 16 days after the LANSCE irradiation with 0.80 GeV protons.

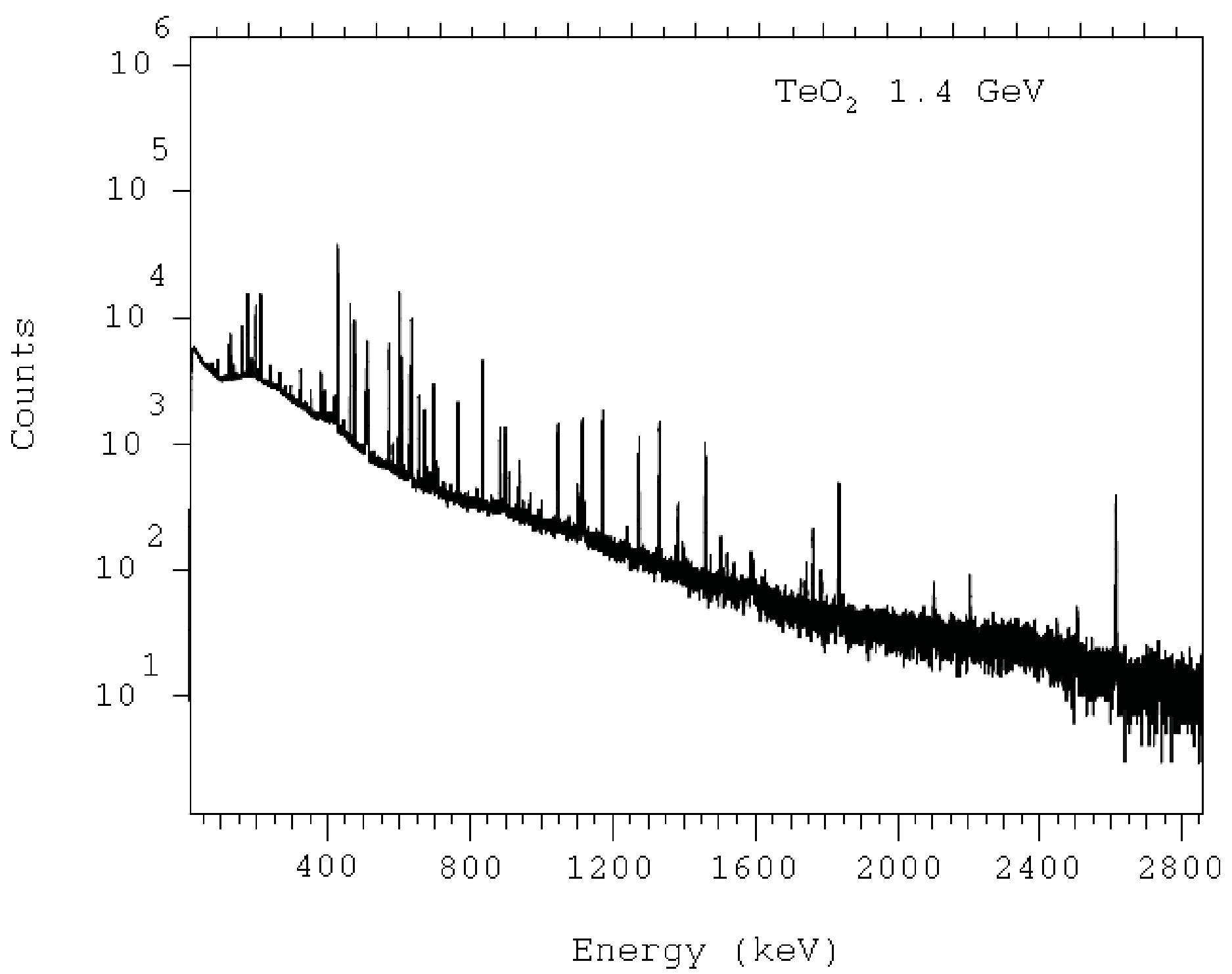

**FIGURE 2(a):** γ-ray spectrum (583178 sec live time) recorded for the sample irradiated at 1.4. GeV 4.58 years after irradiation: complete spectrum

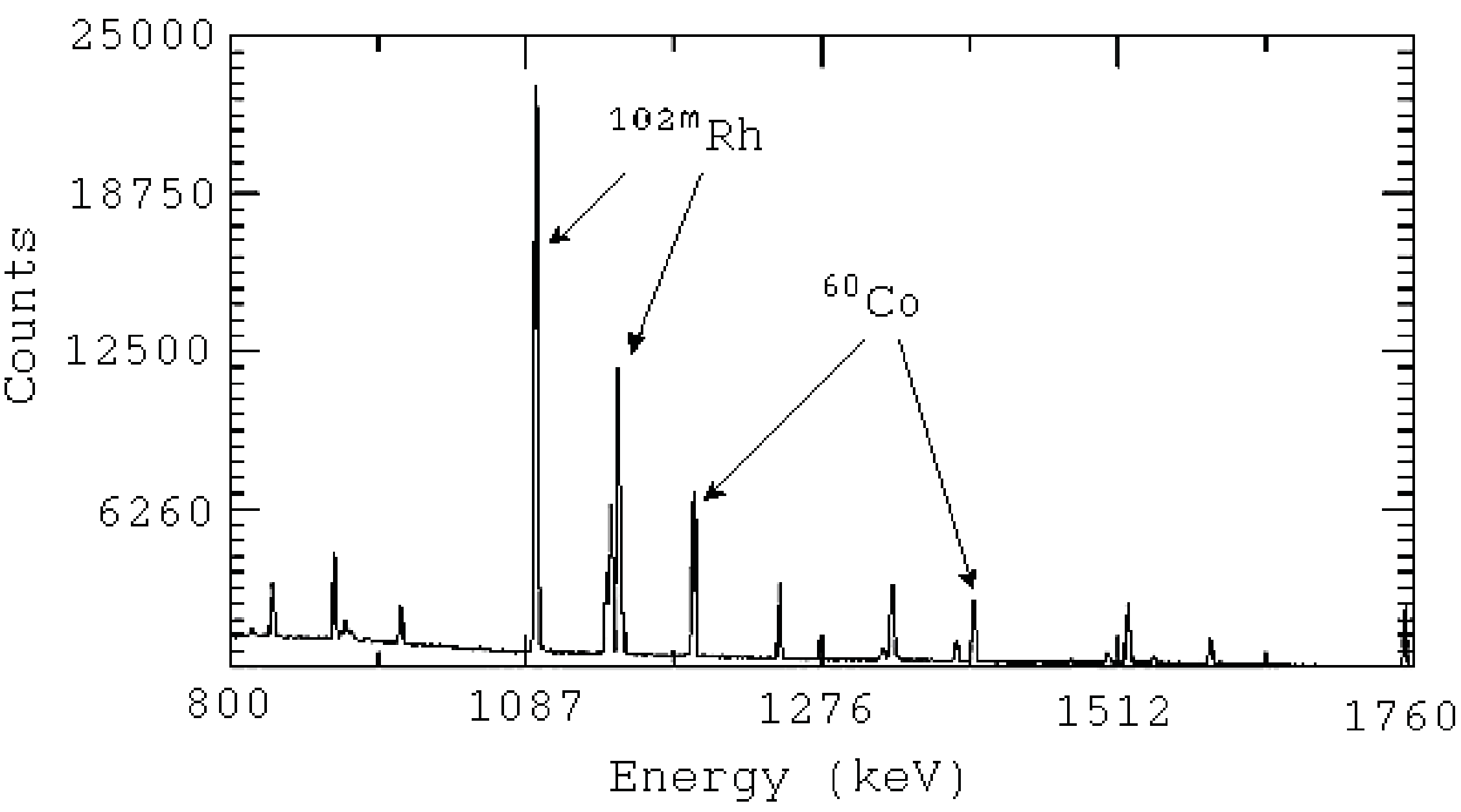

**FIGURE 2(b):** γ-ray spectrum (583178 sec live time) recorded for the sample irradiated at 1.4. GeV 4.58 years after irradiation: peaks of interest.

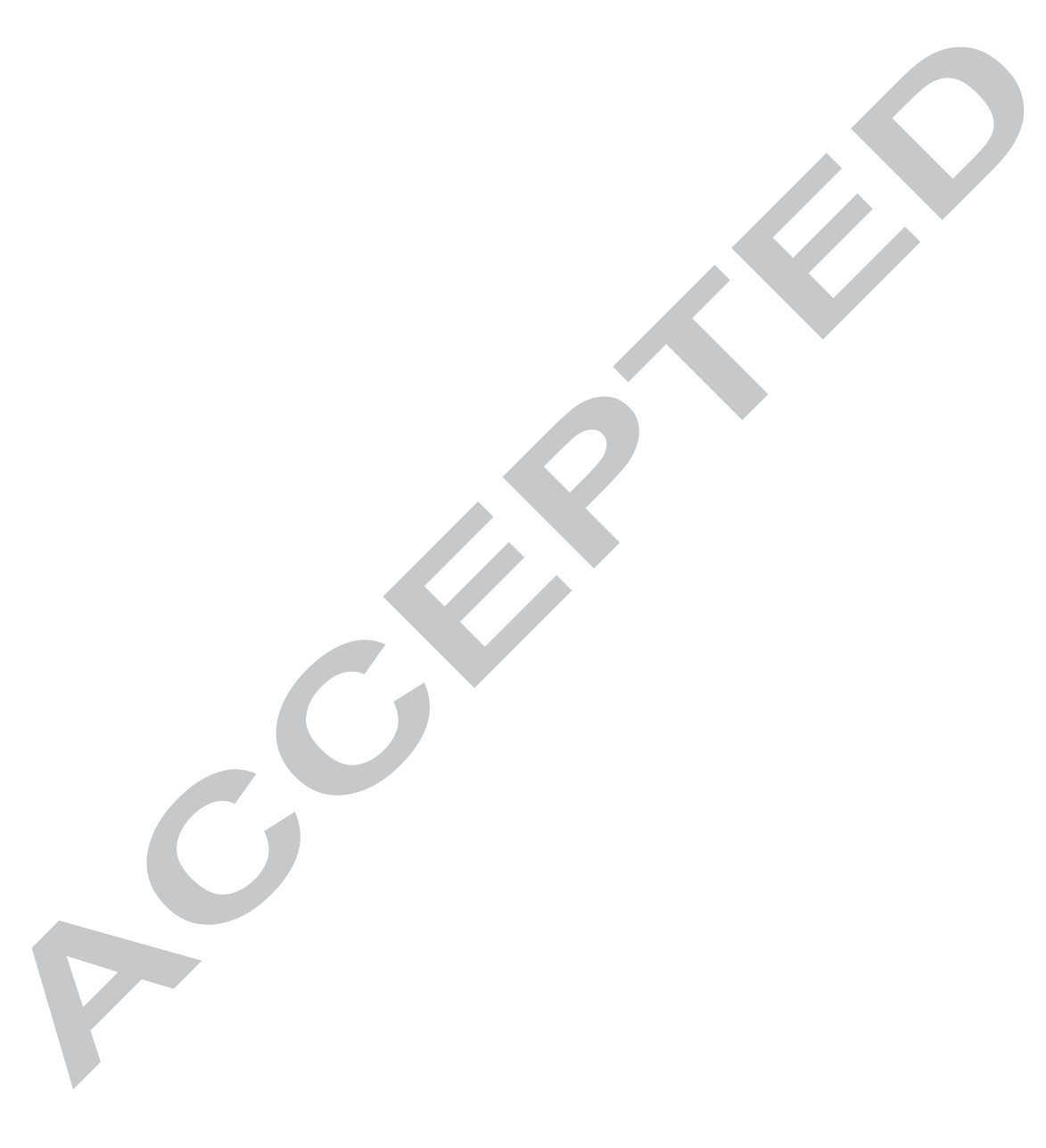

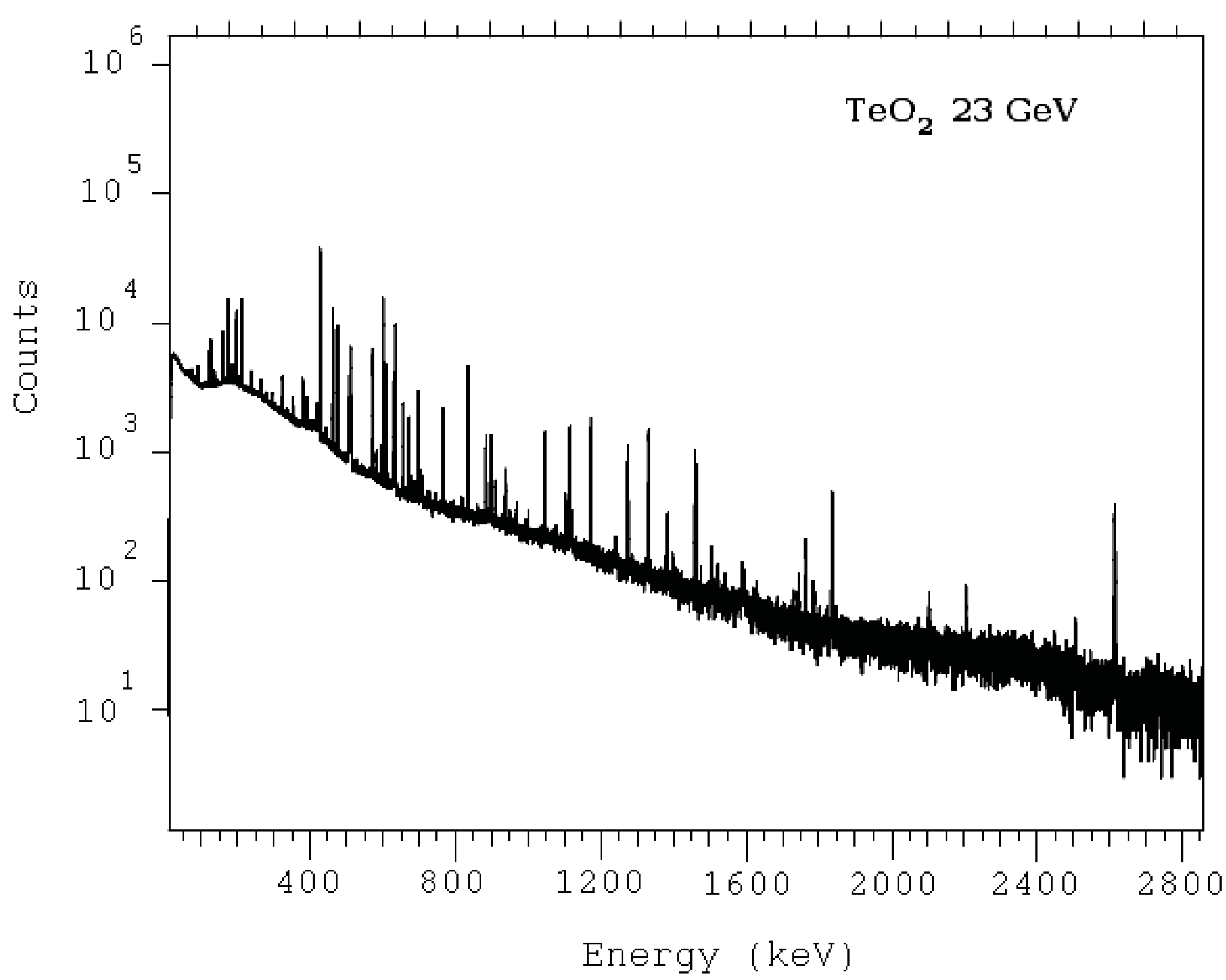

**FIGURE 3(a):** γ-ray spectrum (861228 sec live time) recorded for the sample irradiated at 23 GeV 2.83 years after irradiation: complete spectrum;

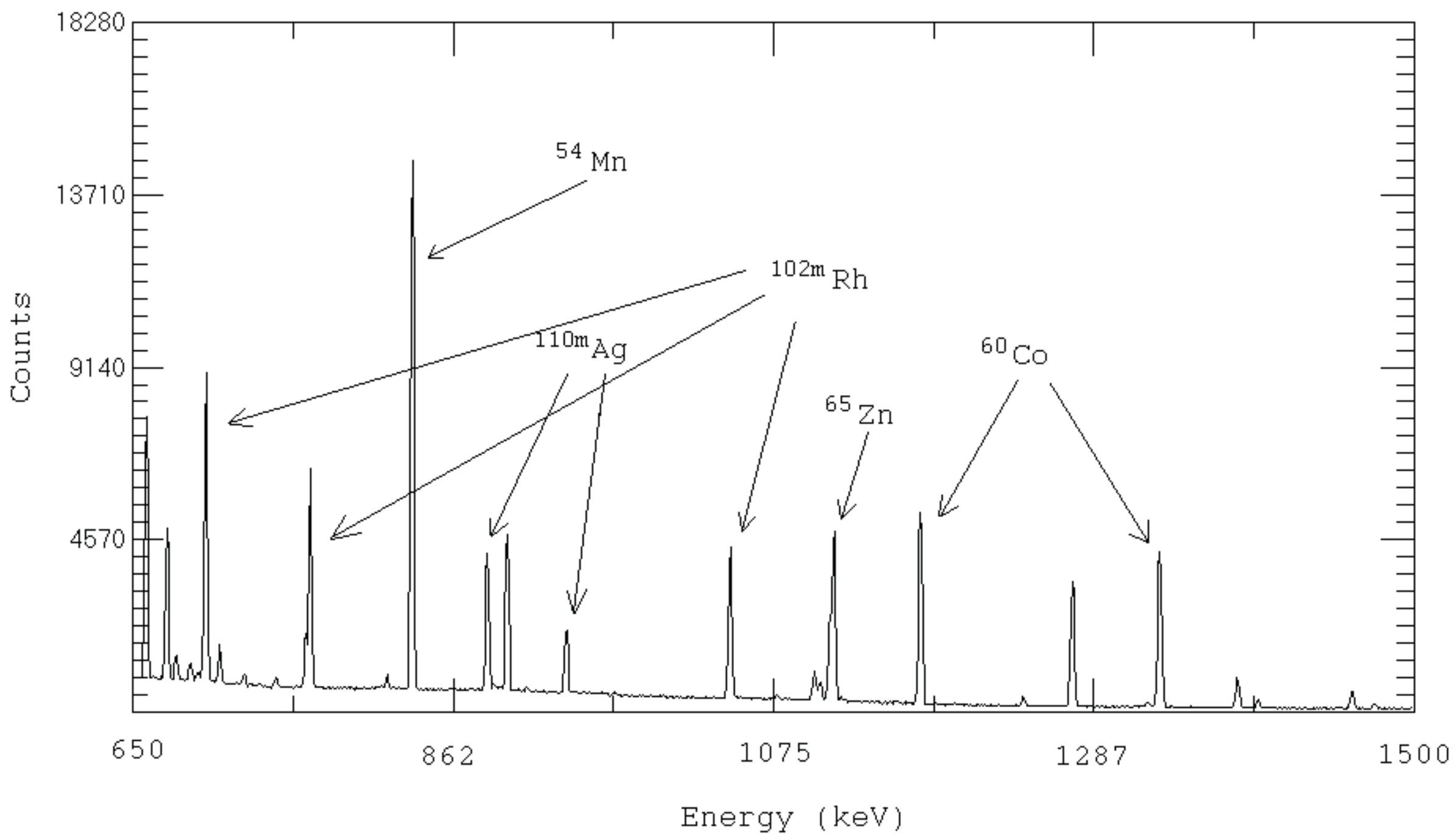

**FIGURE 3(b):** γ-ray spectrum (861228 sec live time) recorded for the sample irradiated at 23 GeV 2.83 years after irradiation: peaks of interest

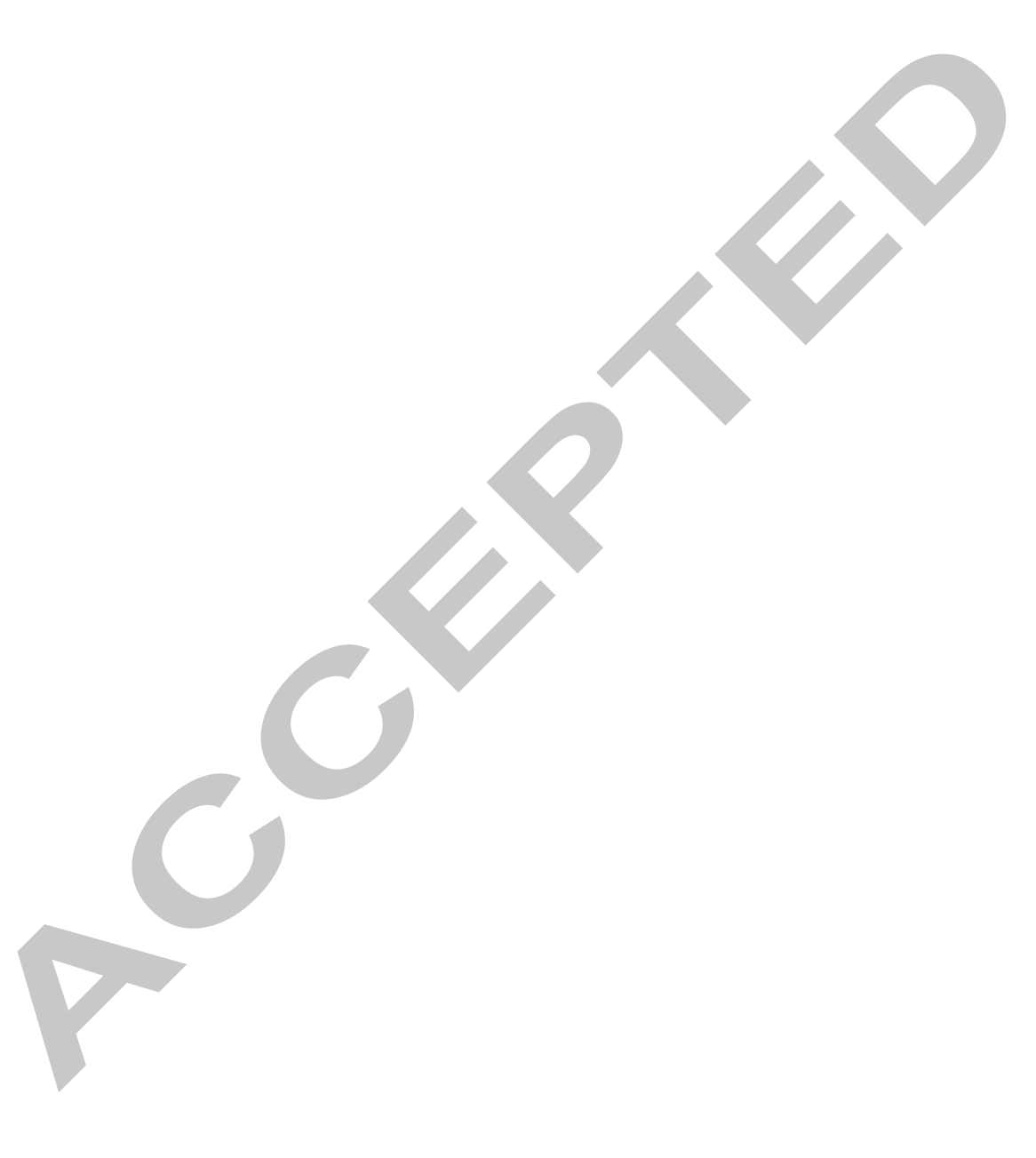